\documentclass[11pt,a4paper,showpacs,aps]{revtex4}
\usepackage[latin1]{inputenc}
\usepackage{amssymb}
\usepackage{amsmath}
\usepackage{graphicx}
\usepackage{fancyhdr}

\begin{document}

\title{Dynamical breaking of gauge symmetry in supersymmetric quantum electrodynamics in three-dimensional spacetime}

\author{A.~C.~Lehum}
\email{lehum@fma.if.usp.br}
\affiliation{Instituto de F\'{\i}sica, Universidade de S\~{a}o Paulo\\
 Caixa Postal 66318, 05315-970, S\~{a}o Paulo - SP, Brazil}


\begin{abstract}

The dynamical breaking of gauge symmetry in the supersymmetric quantum electrodynamics in three-dimensional spacetime is studied at two-loop approximation. At this level, the effective superpotential is evaluated in a supersymmetric phase. At one-loop order, we observe a generation of the Chern-Simons term due to a parity violating term present in the classical action. At two-loop order, the scalar background superfield acquires a nonvanishing vacuum expectation value, generating a mass term $A^{\alpha}A_{\alpha}$ through Coleman-Weinberg mechanism. It is observed that the mass of gauge superfield is predominantly an effect of the topological Chern-Simons term.

\end{abstract}

\pacs{11.30.Pb, 11.30.Qc, 11.15.-q}

\maketitle

\section{Introduction}

The dynamical generation of mass, first proposed by Coleman and Weinberg~\cite{Coleman:1973jx}, is an interesting phenomenon that provides spontaneous symmetry breaking through radiative corrections in the sense of perturbation theory. In four-dimensional spacetime, the gauge symmetry of the scalar quantum electrodynamics with a quartic self-interaction $\lambda\phi^4$ is dynamically broken generating masses to the scalar and gauge fields already at one-loop order. However, for models defined in three-dimensional spacetime, the breaking of gauge symmetry only occurs after two-loop corrections to the effective potential~\cite{Tan:1996kz,Tan:1997ew,Dias:2003pw}.

Supersymmetric extensions of the usual models should share the same properties like the gauge symmetry breaking at two-loop order. In fact, the supersymmetric quantum electrodynamics in $D=2+1$ (SQED$_3$) was studied at one loop showing that no breaking occurs up to this order~\cite{Burgess:1983nu}. Recently, the $3D$ Wess-Zumino model was shown to exhibit a consistent dynamical generation of mass at two-loop approximation due to a dynamical supersymmetry and discrete symmetry breaking~\cite{Lehum:2008vn}. Therefore, apparently a conspiracy in $3D$ models reveals the dynamical generation of mass at two-loop order. The main goal of the present work is to study it in the case of SQED$_3$. To do this, the tadpole equation of the model is studied and a K\"{a}hlerian-like effective potential~\cite{BK0,WZ} (where no derivatives applied in external superfields are considered) is obtained through the tadpole method~\cite{Weinberg:1973ua,Miller:1983fe,Miller:1983ri} in a phase where supersymmetry is manifest.

\section{Effective superpotential in SQED$_3$}

The starting point is to define the SQED$_3$ action
\begin{eqnarray}\label{deq1}
S=\int{d^5z}\Big{\{}\frac{1}{2}W^{\alpha}W_{\alpha}
-\frac{1}{2}\overline{\nabla^{\alpha}\Phi}\nabla_{\alpha}\Phi +M\bar\Phi\Phi\Big{\}},
\end{eqnarray}

\noindent
where  $W^{\alpha}=(1/2)D^{\beta}D^{\alpha}A_{\beta}$ is the Maxwell's field strength and $\nabla^{\alpha}=(D^{\alpha}-ieA^{\alpha})$ is the supercovariant derivative. The conventions and notations are adopted to be as in~\cite{Gates:1983nr}. 

The action above, Eq.(\ref{deq1}), is invariant under the following infinitesimal gauge transformations:
\begin{eqnarray}\label{deq2}
\bar\Phi\longrightarrow \bar\Phi^{\prime}=\bar\Phi(1-ieK),\nonumber\\
\Phi\longrightarrow \Phi^{\prime}=(1+ieK)\Phi,\\
A_{\alpha}\longrightarrow A_{\alpha}^{\prime}=A_{\alpha}+D_{\alpha}K,\nonumber
\end{eqnarray}

\noindent
where $K=K(x,\theta)$ is a real scalar superfield.

This model does not reveal a spontaneous generation of mass to the gauge superfield at tree level like happens in Ref.\cite{Lehum:2007nf}, but we can consider that the superfield $ \Phi $ ($\bar\Phi$) acquire a constant real vacuum expectation value (VEV)  $\varphi/\sqrt{2}$ and study a possible dynamical breaking of the gauge symmetry of the model, where $\varphi$ is interpreted as a classical background superfield. Let us emphasize that no derivatives over $\varphi$ will be considered, in the spirit of the K\"{a}hlerian effective potential~\cite{BK0,WZ}. This assumption allows us to write the superfield $\Phi$ ($\bar\Phi$) as
\begin{eqnarray}\label{deq5}
\bar\Phi=\frac{1}{\sqrt{2}}\Big(\Phi_1+\varphi-i\Phi_2\Big),\nonumber\\
\Phi=\frac{1}{\sqrt{2}}\Big(\Phi_1+\varphi+i\Phi_2\Big),
\end{eqnarray}

\noindent
where $\Phi_1$ and $\Phi_2$ are real scalar superfields with vanishing VEV those possess the following infinitesimal gauge transformations:
\begin{eqnarray}\label{deq5a}
\Phi_2 &\longrightarrow & \Phi_2^{\prime}=\Phi_2+eK(\Phi_1+\varphi),\nonumber\\
\Phi_1 &\longrightarrow & \Phi_1^{\prime}=\Phi_1-eK\Phi_2.
\end{eqnarray}

Thus, the action in terms of the real quantum superfields with vanishing VEV is given by
\begin{eqnarray}\label{deq6}
S&=&\int{d^5z}\Big{\{}\frac{1}{2}W^{\alpha}W_{\alpha}-\frac{e^2\varphi^2}{4}A^{\alpha}A_{\alpha}
+\frac{1}{2}\Phi_1 (D^2+M)\Phi_1 \nonumber\\
&&+\frac{1}{2}\Phi_2 (D^2+M)\Phi_2
+\frac{e}{2}\left( D^{\alpha}\Phi_2 A_{\alpha}\Phi_1
-D^{\alpha}\Phi_1 A_{\alpha}\Phi_2\right)\nonumber\\
&&-\frac{e^2}{2}(\Phi_1^2+\Phi_2^2)A^2
- e^2\varphi\Phi_1 A^2 +M\varphi\Phi_1
+\frac{e\varphi}{2}D^{\alpha}A_{\alpha}\Phi_2
\Big{\}}.
\end{eqnarray}

In order to quantize and eliminate the mixing between the superfields $A^{\alpha}$ and $\Phi_2$ that appears 
in the last term of Eq.(\ref{deq6}), we will use an $R_{\xi}$ gauge fixing plus the corresponding Faddeev-Popov term, introduced through the action
\begin{eqnarray}\label{deq6a}
S_{GF+FP}&=&\int{d^5z}\Big[-\frac{1}{4\alpha}
(D^{\alpha}A_{\alpha}+\alpha e\varphi\Phi_2)^2
+c^{\prime}D^2c+\frac{\alpha}{2}e^2\varphi^2c^{\prime}c+\frac{\alpha}{2}{e^2\varphi}c^{\prime}\Phi_1 c\Big].
\end{eqnarray}

Therefore, the propagators obtained from Eq.(\ref{deq6}+\ref{deq6a}) are given by
\begin{eqnarray}
&&\Delta_{\alpha\beta}(k,\theta,\theta')=-\frac{i}{2}\frac{D^2}{k^2}\left[\frac{D_{\beta}D_{\alpha}}{(k^2+e^2\varphi^2/2)}
+\frac{\alpha}{2}\frac{(D^2-\alpha e^2\varphi^2/2)D_{\alpha}D_{\beta}}{(k^2+\alpha^2e^4\varphi^4/4)}\right]
\delta^{(2)}(\theta-\theta'),\label{deq6b}\\
&&\Delta_{j}(k,\theta,\theta')=-i\frac{(D^2-M)}{k^2+M^2}
\delta^{(2)}(\theta-\theta'),\label{deq6c}\\
&&\Delta_{ghost}(k,\theta,\theta')=\frac{i}{2}\frac{(D^2-M_c)}{k^2+M_c^2}
\delta^{(2)}(\theta-\theta'),\label{deq6d}
\end{eqnarray}

\noindent
where the index $j=1,2$ is related with the indexes of the real superfields $\Phi_1$ and $\Phi_2$, and $M_c=\alpha e^2\varphi^2/2$.   

The minimum of the classical superpotential constraint $\varphi$ is zero. Now, let us evaluate radiative corrections to the classical superpotential using the tadpole method. To understand how we can do that, let us remind that the effective action $\Gamma[\phi_{cl}]$ and the 1PI functions are related, in particular with the tadpole diagrams, by
\begin{eqnarray}\label{deq6e}
\frac{\delta\Gamma[\phi_{cl}]}{\delta\phi_{cl}}=-i\langle\phi(x)\rangle_{1PI},
\end{eqnarray}

\noindent
therefore, integrating the above expression for zero external momentum, the effective potential can be determined from tadpole diagrams as
\begin{eqnarray}\label{deq6f}
-(VT)~V_{eff}=\Gamma[\phi_{cl}]=-i\int\!{d\phi_{cl}}\langle\phi(x)\rangle_{1PI},
\end{eqnarray}

\noindent
where $(VT)$ is the volume of spacetime, and $\phi_{cl}$ is a constant background field.

\begin{figure}[ht]
 \begin{center}
\includegraphics[]{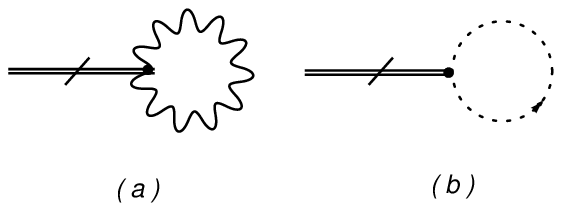}
  \end{center}
\caption{\em One-loop contributions to the tadpole equation. Cut double lines represent the external $\Phi_1$ superfield, wave lines represent the gauge superfield propagator, and dashed lines represent the ghost superfield propagator.} \label{gap1l}
\end{figure}

The one-loop corrections to the tadpole equation are shown in Fig.\ref{gap1l}. The diagram \ref{gap1l}$(a)$ is given by
\begin{eqnarray}\label{deq6g}
\Gamma_{\ref{gap1l}(a)}&=&\frac{e^2}{4}\varphi\int\!\frac{d^3p}{(2\pi)^3}d^2\theta~(2\pi)^3\delta^{(3)}(p)\Phi_1(p)\nonumber\\
&&\int\!\frac{d^3k}{(2\pi)^3}\left[\frac{2}{k^2+e^2\varphi^2/2}
-\alpha\frac{k^2+(\alpha/2) e^2\varphi^2 D^2}{k^2(k^2+\alpha^2e^4\varphi^4/4)}D^2\right]\delta^{(2)}(\theta-\theta),
\end{eqnarray}

\noindent
where the first and the last terms are vanishing because $\delta^2(\theta-\theta)=0$. So, the $\Gamma_{\ref{gap1l}(a)}$ contribution is 
\begin{eqnarray}\label{deq6h}
\Gamma_{\ref{gap1l}(a)}=-\alpha\frac{e^2}{4}\varphi\int\!\frac{d^3p}{(2\pi)^3}d^2\theta~(2\pi)^3\delta^{(3)}(p)\Phi_1(p)
\int\!\frac{d^3k}{(2\pi)^3}\frac{1}{k^2+\alpha^2e^4\varphi^4/4},
\end{eqnarray}

\noindent
where we use the properties $D^2\delta^{(2)}(\theta-\theta)=1$ and $[D^2(k)]^2=-k^2$.

The contribution of \ref{gap1l}$(b)$, the diagram involving the superpropagator of ghosts, is
\begin{eqnarray}\label{deq6i}
\Gamma_{\ref{gap1l}(a)}=\alpha\frac{e^2}{4}\varphi\int\!\frac{d^3p}{(2\pi)^3}d^2\theta~(2\pi)^3\delta^{(3)}(p)\Phi_1(p)
\int\!\frac{d^3k}{(2\pi)^3}\frac{1}{k^2+\alpha^2e^4\varphi^4/4},
\end{eqnarray}

\noindent
that is exactly the "minus" gauge superfield contribution, cancelling the one-loop tadpole equation. 

Even so the classical action does not possess a Chern-Simons term, it can be generated by quantum corrections at one-loop order. This happens because in the classical action there is a parity violating term, the mass of matter superfield. The corresponding mathematical expression can be cast as
\begin{eqnarray}\label{deq6j}
\Gamma_{AA}&=& e^2 \int\!\frac{d^3p}{(2\pi)^3}d^2\theta~f(p,M)\Big{\{}W^{\alpha}W_{\alpha}-MA^{\alpha}W_{\alpha}\Big{\}}~,
\end{eqnarray}

\noindent
where $f(p,M)=\int\frac{d^3k}{(2\pi)^3}\{(k^2+M^2)[(k-p)^2+M^2]\}^{-1}$. It is important to note that the order of generated Chern-Simons term is $e^2$, and no correction at two or more loops is expected~\cite{Coleman:1985zi}. It is well known that Chern-Simons term provides a gauge invariant topological mass to the Maxwell's theory, and this term will be important to analyze the mass of the gauge superfield at two-loop order.

Since the one-loop tadpole equation is vanishing, let us continue searching for a possible Coleman-Weinberg mechanism at two-loop order.

\begin{figure}[ht]
 \begin{center}
\includegraphics[]{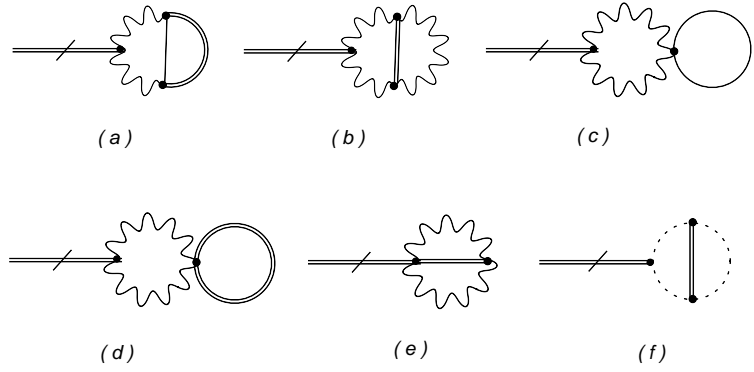}
  \end{center}
\caption{\em Two-loop contributions to the tadpole equation. Double lines represent the $\Phi_1$ superfield propagator and  single lines the $\Phi_2$ superfield propagator.} \label{dsb1}
\end{figure}

The diagram Fig.\ref{dsb1}(f) is vanishing, while the other diagrams, apart from a $\int\!\frac{d^3p}{(2\pi)^3}d^2\theta~(2\pi)^3\delta^{(3)}(p)\Phi_1(p)$ common term, are given by
\begin{eqnarray}\label{deq7}
\Gamma_{\ref{dsb1}(a)}&=&-i\frac{e^4\varphi}{12}~
\int\frac{d^3k}{(2\pi)^3}\frac{d^3q}{(2\pi)^3}\Big\{\frac{M}{(k^2+M_A^2)^2[(k+q)^2+M^2](q^2+M_A^2)}\nonumber\\
&&-\frac{\alpha^3}{4}e^2\varphi^2\frac{12(q^2+M^2)+6(k^2+k.q)}{(k^2+M_c^2)^2[(k+q)^2+M^2](q^2+M^2)}\Big\}\nonumber\\
&=&\frac{i}{384\pi^2}\frac{e^4M\varphi}{M_A(2M+M_A)}+\mathrm{gauge~dependent~part},
\end{eqnarray}
\begin{eqnarray}\label{deq7a}
\Gamma_{\ref{dsb1}(b)}&=&-i\frac{e^6\varphi^3}{12}~
\int\frac{d^3k}{(2\pi)^3}\frac{d^3q}{(2\pi)^3}\Big\{\frac{M~k.q}{k^2(k^2+M_A^2)^2[(k+q)^2+M^2]q^2(q^2+M_A^2)}\nonumber\\
&&+\frac{\alpha^6}{128}e^{8}\varphi^8\frac{k.q}{k^2(k^2+M_c^2)^2[(k+q)^2+M^2]q^2(q^2+M_c^2)}\nonumber\\
&&+\frac{\alpha^5}{96}e^6\varphi^6\frac{4q^2-3k.q+k^2}{(k^2+M_c^2)^2[(k+q)^2+M^2]q^2(q^2+M_c^2)}\nonumber\\
&&+\frac{\alpha^4}{32}e^4M\varphi^4\frac{2q^2~k^2-k^2~k.q+e^2\varphi^2~k.q}{k^2(k^2+M_c^2)^2[(k+q)^2+M^2]q^2(q^2+M_c^2)}\nonumber\\
&&-\frac{\alpha^3}{16}\varphi^2\frac{k^2~q^2-q^2~k.q-2e^2\varphi^2(k.q+q^2)}{(k^2+M_c^2)^2[(k+q)^2+M^2]q^2(q^2+M_c^2)}\nonumber\\
&&+\frac{\alpha^2}{8}M\varphi^2\frac{k^2~k.q-e^2\varphi^2~k.q}{k^2(k^2+M_c^2)^2[(k+q)^2+M^2]q^2(q^2+M_c^2)}\nonumber\\
&&+\frac{\alpha}{4}\varphi^2\frac{k^2-k.q}{k^2(k^2+M_c^2)^2[(k+q)^2+M^2](q^2+M_c^2)}\Big\}\nonumber\\
&=&\frac{i}{384\pi^2}\frac{e^6M\varphi^3}{M_A^6(M+2M_A)}\Big{\{}M^2(M+2M_A)\ln\left(\frac{M}{\mu}\right)
+(M+M_A)M_A^2\nonumber\\
&&-(2M^3-4M^2M_A-MM_A^2-2M_A^3)\ln\left(\frac{M+M_A}{\mu}\right)\nonumber\\
&&+(M+M_A)(M^2+MM_A-2M_A^2)\ln\left(\frac{M+2M_A}{\mu}\right)\Big{\}}+\mathrm{gauge~dependent~part},
\end{eqnarray}
\begin{eqnarray}\label{deq7bb}
\Gamma_{\ref{dsb1}(c+d)}&=&-i\frac{\alpha^3}{4}{e^6\varphi^3}~
\int\frac{d^3k}{(2\pi)^3}\frac{d^3q}{(2\pi)^3}\frac{1}{(k^2+M_c^2)^2(q^2+M^2)},
\end{eqnarray}
\begin{eqnarray}\label{deq7b}
\Gamma_{\ref{dsb1}(e)}&=&-i\frac{e^4\varphi}{8}~
\int\frac{d^3k}{(2\pi)^3}\frac{d^3q}{(2\pi)^3}\Big\{\frac{M~k.q}{k^2(k^2+M_A^2)[(k+q)^2+M^2]q^2(q^2+M_A^2)}\nonumber\\
&&+\frac{\alpha}{2}\frac{(k^2+q^2)k.q+2k^2~q^2}{k^2(k^2+M_A^2)[(k+q)^2+M^2]q^2(q^2+M_A^2)}\nonumber\\
&&+\frac{\alpha^2}{4}\frac{M}{(k^2+M_c^2)[(k+q)^2+M^2](q^2+M_A^2)}\nonumber\\
&&+\frac{\alpha^3}{4}e^2\varphi^2\frac{k^2~q^2-2(k^2+q^2)k.q}{k^2(k^2+M_c^2)[(k+q)^2+M^2]q^2(q^2+M_A^2)}\nonumber\\
&&+\frac{\alpha^4}{16}e^4\varphi^4M\frac{k.q}{k^2(k^2+M_c^2)[(k+q)^2+M^2]q^2(q^2+M_A^2)}\Big\}\nonumber\\
&=&\frac{i}{128\pi^2}\frac{e^4M\varphi}{M_A^4}\Big{\{} M_A^2
+2(M_A^2-M^2)\ln\left(\frac{M+M_A}{\mu}\right)\nonumber\\
&&+(M^2-2M_A^2)\ln\left(\frac{M+2M_A}{\mu}\right)+M^2\ln\left(\frac{M}{\mu}\right)\Big{\}}+\mathrm{gauge~dependent~part},
\end{eqnarray}

\noindent
where $M_A^2=e^2\varphi^2/2$ and $M_c^2=\alpha^2e^4\varphi^4/4$. To perform the integrals, regularization by dimensional reduction~\cite{Siegel:1979wq} was used, and therefore $\mu$ is the mass scale introduced by this regularization scheme. The two-loop D-algebra calculations were made with the help of a MATHEMATICA$^{\copyright}$ package~\cite{Ferrari:2007sc}.

Adding the tree and two-loop contributions, we obtain the following tadpole equation:
\begin{eqnarray}\label{deq8}
\Gamma_{\ref{dsb1}}(\varphi)&=& \frac{i}{96\pi^2}\frac{M}{\varphi^3(M+\sqrt{2}e\varphi)(4M+\sqrt{2}e\varphi)}\Big{\{}
192\pi^2e^2\varphi^6 + 480\pi^2\sqrt{2}e M \varphi^5\nonumber\\
&&+\varphi^4\left[5e^4+384\pi^2M^2+8e^4\ln\left(1+\frac{e\varphi}{\sqrt{2}M}\right)-8e^4\ln\left(1+\frac{\sqrt{2}e\varphi}{M}\right)\right]\nonumber\\
&&+\varphi^3\sqrt{2}\left[11e^3M+20e^3M\ln\left(1+\frac{e\varphi}{\sqrt{2}M}\right)-20e^3M\ln\left(1+\frac{\sqrt{2}e\varphi}{M}\right)\right]\nonumber\\
&&+\varphi^2\left[10e^2M^2-4e^2M^2\ln\left(1+\frac{e\varphi}{\sqrt{2}M}\right)-6e^2M^2\ln\left(1+\frac{\sqrt{2}e\varphi}{M}\right)\right]\nonumber\\
&&+\varphi\left[-50eM^3\ln\left(1+\frac{e\varphi}{\sqrt{2}M}\right)+25eM^3\ln\left(1+\frac{\sqrt{2}e\varphi}{M}\right)\right]\nonumber\\
&&+20M^4\left[\ln\left(1+\frac{\sqrt{2}e\varphi}{M}\right)-2\ln\left(1+\frac{e\varphi}{\sqrt{2}M}\right)\right]\Big{\}}+\mathrm{gauge~dependent~part}.
\end{eqnarray}

To evaluate the effective superpotential, it is convenient to expand the tadpole equation (\ref{deq8}) for $e\ll 1$,
\begin{eqnarray}\label{deq9}
\Gamma_{\ref{dsb1}}(\varphi)&=& i\Big\{ \frac{e^3}{384\sqrt{2}\pi^2}+M\varphi-\frac{7e^4}{1536\pi^2M}\varphi
+\frac{83e^5}{9216\sqrt{2}\pi^2M^2}\varphi^2\nonumber\\
&&-\frac{275e^6}{36864\pi^2M^3}\varphi^3+\mathcal{O}(e^7)\Big\}+\mathrm{gauge~dependent~part}.
\end{eqnarray}

The effective superpotential can be obtained integrating Eq.(\ref{deq9}) over $\varphi$ as in Eq.(\ref{deq6f}),  resulting in
\begin{eqnarray}\label{deq11}
V_{eff}(\varphi_{cl})&=&-\Big\{ \frac{e^3}{384\sqrt{2}\pi^2}\varphi_{cl}+\left(\frac{M}{2}-\frac{7e^4}{3072\pi^2M}\right)\varphi_{cl}^2+\frac{83e^5}{27648\sqrt{2}\pi^2M^2}\varphi_{cl}^3\nonumber\\
&&-\frac{275e^6}{147456\pi^2M^3}\varphi_{cl}^4+\mathcal{O}(e^7)\Big\}+\mathrm{gauge~dependent~part}.
\end{eqnarray}

\noindent
Notice that the effective superpotential is gauge dependent, agreeing with a previous result obtained by Jackiw in the scalar electrodynamics~\cite{Jackiw:1974cv}. Furthermore, the effective superpotential is finite as argued in Ref.\cite{Ferrari:2007mh}. For now, for simplicity and without lost of generality, let us choose $\alpha=0$. 

It is easy to see that $\varphi=0$ is not a solution that vanishes the tadpole equation, i.e., it is not a solution that minimizes the effective superpotential. Therefore, the background superfied $\varphi$ acquired a nonvanishing vacuum expectation value at two-loop order. The value of $\varphi$ that is a solution to the tadpole equation $\Gamma_{\ref{dsb1}}(\varphi)=0$, can be evaluated order by order in the coupling constant $e$. Considering the tadpole equation up to the order of $\mathcal{O}(e^4)$, the minimum of the effective superpotential is given by
\begin{eqnarray}\label{deq10}
\varphi_0=-\frac{e^3}{384\sqrt{2}\pi^2M}+\mathcal{O}(e^7),
\end{eqnarray}

\noindent
where $\varphi_0$ is the value of the background superfield $\varphi$ in the minimum of the effective superpotential. This nonvanishing value of $\varphi_0$ is a manifestation of the Coleman-Weinberg mechanism. The minimum of the classical superpotential $\varphi_0=0$ becomes a local maximum, dislocating the minimum by $\varphi_0=-\frac{e^3}{384\sqrt{2}\pi^2M}+\mathcal{O}(e^7)$. 

The gauge superfield $A^{\alpha}$ acquires a mass term, $M_A^2=e^2\varphi_0^2/2$, due to the spontaneous gauge symmetry breaking. But, at one-loop order we showed that a Chern-Simons topological mass was induced by quantum corrections. So, what about gauge superfield mass?

After two-loop corrections, the quadratic part of the gauge superfield effective action possesses the form
\begin{eqnarray}\label{max1}
S_{eff}&=&\int{\frac{d^3p}{(2\pi)^3}}d^2\theta~\Big\{\frac{f_1}{2}W^{\alpha}W_{\alpha}+\frac{f_2}{2}A^{\alpha}W_{\alpha}+\frac{f_3}{2}A^{\alpha}A_{\alpha}-\frac{1}{4\alpha}
(D^{\alpha}A_{\alpha})^2+\mathrm{non~local~corrections}\Big\}\nonumber\\
&=&\int{\frac{d^3p}{(2\pi)^3}}d^2\theta~\Big\{\frac{1}{2}A_{\gamma}\mathcal{O}^{\gamma\beta}A_{\beta}+\mathrm{non~local~corrections}\Big\},
\end{eqnarray}

\noindent 
where $f_1$, $f_2$, and $f_3$ are the local corrections to the Maxwell, Chern-Simons and mass term, respectively.

The inverse of the $\mathcal{O}^{\gamma\beta}$ operator is given by
\begin{eqnarray}\label{max2}
(\mathcal{O}^{-1})^{\beta\gamma}(p,\theta,\theta')&=&-\frac{i}{2}\frac{D^2}{p^2}\Big\{a_1
\Big[\frac{(D^2-a_2)}{(p^2+a_2^2)}-\frac{(D^2-a_4)}{(p^2+a_3^2)}\Big]D_{\gamma}D_{\beta}\nonumber\\
&&+\alpha\frac{(D^2+\alpha f_3)D_{\beta}D_{\gamma}}{(p^2+\alpha^2 f_3^2)}\Big\}
\delta^{(2)}(\theta-\theta')~,
\end{eqnarray}

\noindent
where
\begin{eqnarray}\label{max2a}
a_1&=&\frac{1}{\sqrt{f_2^2-4f_1f_3}}\nonumber\\
a_2&=&\frac{f_2}{2f_1}\left(1+\sqrt{1-\frac{4f_1f_3}{f_2}}\right)\\
a_3&=&\frac{f_2}{2f_1}\left(1-\sqrt{1-\frac{4f_1f_3}{f_2}}\right).\nonumber
\end{eqnarray}

From the gauge-independent part of the propagator we can identify three poles, one in $p^2=0$ and the others in $p^2=-a_2^2$ and $p^2=-a_3^2$. 

From Eq.(\ref{deq6j}) we can see that the induced Chern-Simons term $f_2$ is of the order of $e^2$, and the correction to the Maxwell's term is $f_1=1+\mathcal{O}(e^2)$. The mass term $A^{\alpha}A_{\alpha}$ is generated at two loop, being of the order of $e^8/M$. Therefore, the massive poles of the gauge superfield propagator is given approximately in 
$p^2\sim -e^4\left(1\pm\sqrt{1-\frac{e^6}{M}}\right)^2$, i.e., $p^2\sim-(e^4+e^{16}/M)\approx -e^4$ and $p^2\sim-e^{16}/M$. The pole $p^2\sim-e^4$ in the present model is much bigger than that generated due to gauge symmetry breaking, being predominantly generated by effects of the topological Chern-Simons term.

\section{Concluding remarks}

In this work the effective superpotential was evaluated at two-loop approximation in the context of the K\"{a}hlerian effective potential, considering only the terms where no derivatives are applied on the constant background superfield. The model was shown to be spontaneously broken at this level, generating a very small mass term to the gauge superfield, smaller than that generated at one-loop order by the Chern-Simons term. Therefore, the mass of the gauge superfield is predominantly an effect of the topological Chern-Simons term. 

Differently from the four-dimensional case, in the three-dimensional spacetime there is no chirality, therefore the general structure of the effective superpotential is different, i.e., in the 4D effective superpotential there always appears a $d^4\theta=d^2\theta d^2\bar\theta$ integration~\cite{Grisaru:1979wc}, while in 3D only a $d^2\theta$ integration appears. Besides this, in three-dimensional supersymmetric quantum electrodynamics all loop corrections are finite, whereas in four dimensions all loop corrections are divergent. Therefore, after renormalization, the 4D effective superpotential depends on an arbitrary scale. 

An interesting point to consider in the future is the possibility of the dynamical supersymmetry breaking of this model, a study that should be in components in the spirit of the work in~\cite{Lehum:2008vn}. Another point to consider is a noncommutative generalization of the present model, because the supersymmetric quantum electrodynamics acquires a structure very similar to a supersymmetric Yang-Mills theory, where self-interacting terms to the gauge superfield must be taken into account.

\vspace{1cm}
{\bf Acknowledgments.} This work was partially supported by Funda\c{c}\~{a}o de Amparo 
\`{a} Pesquisa do Estado de S\~{a}o Paulo (FAPESP) under Project No. 2007/08604-1. The author would like to thank A. Yu Petrov and A. J. da Silva for useful discussions and comments.

\newpage 

\end{document}